\documentclass{aa}  

\usepackage{graphicx}
\usepackage{txfonts}
\usepackage{subfigure}
\usepackage{siunitx}
\begin{document}

   \title{Mira variable stars in the galactic bulge and halo}

   \author{ K.~Neumannov{\'a} \inst{1} 
            }

   \institute{ Department of Theoretical Physics and Astrophysics, Masaryk University,
            Kotl\'a\v{r}sk\'a 2, CZ-611\,37 Brno, Czechia \\
            \email{474150@mail.muni.cz}
         }
   
   \date{2026}

 
  \abstract
   {}
   {This work aimed to study Miras in the bulge and halo of our galaxy, including those in globular clusters.}
   {The AAVSO International Variable Star Index VSX catalogue was used to retrieve Mira variable stars, while the information about globular clusters were taken from the Harris \citep{Harris1976}, McMaster catalogue \citep{33} and Holger Baumgardt \citep{34} table.
Stellar characteristics such as metallicity, $T_{\rm eff}$, parallax and colors were obtained by Gaia EDR3 catalogue, Gaia DR3 VariSumarry, Gaia DR3 catalogue, 2MASS Point Source Catalogue, AllWISE, and ZTF catalogue.
The UPSILoN program, color-color or color-magnitude diagrams were used to verify that our stars are indeed Miras-type variable stars and whether they are O or C type Miras.
To deal with extinction, which affects stars mainly in the galactic bulge, we used IRSA DUST archive, OGLE-III and recalculation to the extiction-free Wesenheit magnitude.}
   {We obtained data for 1608 Miras in the bulge, of which 308 are located in 15 globular clusters. For the stars in the halo, we obtained data for 1954 Miras, of which 4 are located in 3 globular clusters. Miras in the bulge show a bimodal metallicity distribution with peaks at \text{--}0.4 dex and $+$0.5 dex, while Miras in the halo show only one peak at $+$0.5 dex. The period of Miras in the bulge is, on average, 30 days longer than for Miras in the halo, and O-type Miras are more abundant in the bulge than C-type Miras. The dependence of the logarithm of the base 10 period on magnitude in the infrared part of the spectrum gives us a linear dependence, with the dependence inverting when a filter is used in the optical part of the spectrum, because of the dust around the Miras.}
   {}


   \maketitle
%

\section{Introduction}
Mira variables are pulsating variable cool giants located on the asymptotic giant branch (AGB) of the  Hertzsprung–Russell (HR) diagram. Low- to intermediate-mass stars enter the Miras phase on the AGB \citep{Encyclopedia_PST}. They have a high luminosity ($10^4$ ~\(\textup{L}_\odot\)) and a low effective temperature ($T_{\rm eff}$) of up to 4000 K \citep{Encyclopedia_PST}. Because of their already advanced evolutionary stage, during which the star's size increases, Miras have radii up to 2 AU, and the low $T_{\rm eff}$ in their atmospheres can lead to the formation of silicate or graphite dust in the extended circumstellar envelope several stellar radii above the photosphere \citep{Reid2002}. Periods of luminosity variation range from 80 days to 1000 days, where, due to the high period dispersion, some long-period variables (LPVs) include Mira, semiregular and irregular variables \citep{https://doi.org/10.48550/arxiv.1207.4094}.

In addition to their long periods, Miras are characterized by high observed amplitudes of luminosity variations, especially in the optical, where variations reach values of up to 11 mag \citep{Encyclopedia_PST}. In contrast, amplitudes of at most 1 mag are observed in the infrared due to the fact that the infrared amplitudes are smaller because circumstellar dust is more transparent at infrared wavelengths and temperature variations have a weaker effect on the infrared flux \citep{17}.

The most widely accepted theory explaining stellar pulsations in Miras is the $\kappa$-mechanism. This mechanism is well known for pulsating stars in the instability strip, such as RR Lyrae stars and Cepheids. However, in Mira variables, the hydrogen ionization zone plays a dominant role in driving the pulsations, unlike in classical instability-strip pulsators, where the helium ionization zone is primarily responsible. During compression, hydrogen becomes partially ionized, increasing the opacity and temporarily trapping radiation. The accumulated thermal energy raises the internal pressure and drives the outer layers outward. As the atmosphere expands and cools, hydrogen recombines, the opacity decreases, radiation escapes, and gravity causes the atmosphere to contract again. This process repeats itself over and over again \citep{Boyd2021} and \citep{Adams2013}.

If we are interested in qualitative values, during the pulsations near minimum light, the photospheric temperature may decrease to approximately 1900 K, while temperatures of about 1400 K, suitable for dust condensation, are reached at approximately 1.8 stellar radii from the stellar centre \citep{Reid2002}.

Variable stars can be found in all parts of the Universe. Consequently, Mira variables are also observed in globular clusters (GC).
The identification of Mira variables in globular clusters is particularly valuable because cluster ages, distances, and metallicities are generally well constrained. This provides a unique opportunity to investigate how pulsation properties and mass-loss processes depend on the stellar environment and chemical composition.

Mira variables are important astrophysical objects because they represent the final stages of evolution of low- and intermediate-mass stars. Their well-defined period--luminosity relation, particularly in the near-infrared, makes them valuable distance indicators, while their strong mass loss and dust production provide key constraints on late stellar evolution and the chemical enrichment of the interstellar medium. Owing to their high luminosities, Mira variables can also be used to trace the structure and stellar populations of the Milky Way and nearby galaxies.

\section{Data and Methodology}
Color-magnitude diagrams (CMDs) and moving CMDs were used to examine the photometric properties of the selected Mira variables and to compare their distribution with the expected loci of evolved giant stars.

Long-term photometric light curves are essential for analysing Mira variables because their pulsation periods typically range from several months to several years. Therefore, time-series observations covering multiple pulsation cycles are required.

Photometric time-series data were obtained from the Zwicky Transient Facility (ZTF), which performs a wide-field survey of the northern sky in the \textit{g}, \textit{r}, and \textit{i} photometric bands using the 48-inch Samuel Oschin Telescope at Palomar Observatory \citep{Masci2018}. ZTF light curves were accessed through the IRSA/IPAC archive and downloaded for further analysis. The Infrared Science Archive (IRSA), hosted by the Infrared Processing and Analysis Center (IPAC), provides access to photometric and time-series data from numerous ground- and space-based surveys.

With IRSA and IPAC, it is possible to get time series data from several terrestrial and space-based surveys in one place, in different filters. Additional photometric data from other surveys available through IRSA/IPAC were used where necessary.

The AAVSO International Variable Star Index VSX catalogue \citep{2015yCat....102027W} was used to retrieve Mira variable stars, identifying over 28 000 of them in our Galaxy. In addition to identifying Miras with their coordinates, the catalogue also provided us with information on the period of luminosity changes for Miras.

The next step was to select Mira variables according to their Galactic location. Since globular clusters are predominantly found in the Galactic bulge and halo, only Mira variables located in these two Galactic components were considered for further analysis. To determine the boundary of the galactic bulge, we took as its centre the coordinates of Sagittarius~A* -- a massive black hole, with coordinates 17\,h 45\,m 40.0409\,s, \ang{-29;00;28.118} (J2000)  \citep{Reid_2004}, which we transformed into galactic coordinates (galactic longitude -- glon and galactic latitude -- glat) for our purposes.  Following \citet{2017ApJ...849L..24M} and the Galactic bulge morphology presented by \citet{30}, the bulge was defined by Galactic longitudes of 344--360\,deg and 0--14\,deg, and Galactic latitudes of $-15^\circ \le b \le +15^\circ$.

To determine the limit of stars located only in the Galactic halo, were used sources \citep{31}. For the glon, is taken its full range, i.e. 0--360\, deg, because we are interested in both the inner and outer halo and in the whole Galaxy. To reduce contamination from the Galactic disc, only objects with Galactic latitudes $|b|>20^\circ$ were considered halo candidates. Galactic longitude covered the full range of $0^\circ$–$360^\circ$. This range narrowed down the number of Miras to 1954, which are located in the Galactic halo.

\subsection{Mira variable stars in globular clusters}
Only globular clusters with available structural parameters and proper-motion measurements were included, resulting in a final sample of 160 clusters, whose list and coordinates were taken from the catalogue \citep{33} and the proper motion ($\mu$) for these globular clusters was taken from \citep{34}. The proper motion for the stars was obtained from the Gaia DR3 catalogue in Vizier \citep{Eyer2019}, along with the standard error for each star. Similar values of the proper motion of a cluster and a star can be an indication that the star is a member of a globular cluster. This occurs because its members are gravitationally bound to each other, moving collectively in the same direction and at the same speed within space. The total proper motion of both the cluster and each Mira variable was calculated as $\mu = \sqrt{\mu_{\alpha}^{2} + \mu_{\delta}^{2}}$,
where $\mu_\alpha$ and $\mu_\delta$ denote the proper motions in right ascension and declination, respectively. A tolerance of 5$\sigma$  was adopted to minimise the rejection of genuine cluster members while accounting for the uncertainties in the Gaia proper-motion measurements.. This deviation indicates the range in which a given star is likely to be due to measurement inaccuracies.

An initial search radius of $1^\circ$ around each cluster centre was adopted to ensure that potential candidates were not excluded before applying the proper-motion criterion. By satisfying these conditions, 304 Miras were obtained out of the original 1608 stars located in the bulge of the Galaxy. These Miras are distributed in 15 different globular clusters, with the largest number of Miras in Terzan 6, NGC 5453, and Djor 1. In contrast, in the Galactic halo, in which were found 1954 Miras, only 4 of them are located in globular clusters, namely NGC 6093 (2 Miras), Arp 2, and Terzan 8.

We can also check whether a star belongs to a globular cluster using a density profile, where projected stellar density is expected to decrease with increasing distance from the cluster centre. Before looking at individual Miras in globular clusters, it was necessary to analyse the number of Miras as a function of core and half-light radius. The reason was to see if our Miras data are biased by the fact that a given cluster is well resolved or is closer to us, or has a really high star density. 

\subsection{Determination of Stellar Parameters}

Using the final sample of Mira variables, we retrieved additional stellar parameters including metallicity, $T_{\rm eff}$, and parallax. These data were crucial for a better understanding of Miras, and these parameters were used in the subsequent analysis. The parallax for each star was obtained from the Gaia EDR3 catalogue \citep{2021A&A...649A...1G}, while the metallicity and $T_{\rm eff}$ were taken from the Gaia DR3 catalogue \citep{Eyer2019}. At the same time, thanks to the knowledge of the period, we were able to calculate the bolometric absolute magnitude ($M_{\rm bol}$) using the Eq. (\ref{Mbol}). This relation was taken from \citep{Andriantsaralaza2022-xd}. The bolometric magnitude provides an estimate of the total stellar luminosity integrated over all wavelengths. Moreover, $M_{\rm bol}$ is not affected by atmospheric absorption and scattering, giving us a more accurate estimate of the luminosity than a magnitude measured only in the visible part of the spectrum. From this relationship, we can therefore derive the commonly known relationship for luminosity ($L$) in units of $L_{\rm Sun}$, see Eq. (\ref{Lsun}). The value $3.846 \cdot 10^{26}$ is the luminosity of the Sun, and the value $3.0128 \cdot 10^{28}$ is the zero-point luminosity in units of watts. Zero-point luminosity is the luminosity that a star with $M_{\rm bol}$ equal to zero would have and was determined by the IAU in 2015, as was $L_{\rm Sun}$ \citep{2015arXiv151006262M}.

\begin{align}
M_{\rm bol} = -3.31(\log P-2.5)-4.317 \label{Mbol}
\end{align}

\begin{align}
L/L_{\rm \odot}= \frac{3.0128 \cdot 10^{28} \cdot 10^{-0.4 M_{\rm bol}}}{3.846 \cdot 10^{26}} \label{Lsun}
\end{align}

\subsection{Photometric Data and Color Indices}

$T_{\rm eff}$ provided by Gaia DR3 were used to validate temperature estimates derived from photometric color indices. Because Gaia DR3 contains temperature estimates for only approximately one third of the analysed Mira variables \citep{Eyer2019}, the photometric calibration enabled the temperature distribution to be extended to the full sample. The same color indices were subsequently employed in the analysis of color-magnitude and color-color relations.

 Since Miras are rather cooler stars emitting mostly at longer wavelengths, we focused on filters corresponding to this. We used catalogues such as the 2MASS Point Source Catalogue for $J$ and $K$ band \citep{2000yCat.2241....0S}, The AllWISE data release ($W1$ -- 3.35$\rm \mu m$, $W2$ -- 4.6$\rm \mu m$, $W3$ -- 11.6$\rm \mu m$, $W4$ -- 22.1$\rm \mu m$) from \cite{2014yCat.2328....0C}, Gaia DR3 ($BP$ -- blue photometer, $RP$ -- red photometer and $G$ -- broad white light), \citep{2022yCat.1355....0G}. Among other things, we used the Gaia DR3 VariSumarry catalogue \citep{2023A&A...674A..13E}, which provides the minimum and maximum magnitudes for the $G$, $BP$ and $RP$ bands. This allows to observe how the color, and therefore the $T_{\rm eff}$, of a variable star changes during the period at which the brightness changes.

The last source from which we obtained information about the colors was the ZTF catalogue. The data, which included among others modified Julian Date (mjd), magnitude (mag) and filter, were downloaded separately for halo and bulge, with a tolerance of 10 arcsec from our specified coordinates. A matching radius of 10 arcsec was adopted to account for the positional uncertainties of the catalogues. For Miras, which are located in the galactic bulge, we extracted data for 207 stars from the 1608 stars we entered in the ZTF catalogue, 25 of which are located in clusters. For 1954 Miras, which are located in a halo, we obtained data for 855 stars, of which 2 are located in clusters. The ZTF catalogue has an interactive interface that allows us to visualize the data we are looking for and to see where they are located in the Galaxy. This interface therefore serves as a quick check to confirm whether we have indeed selected data from the Galactic bulge/halo. Halo Mira variables were more frequently observed by ZTF than bulge objects, but we also retrieved data for the halo in the $g$, $r$, and $i$ filters, while for the bulge most of the data was only in the $g$ and $r$ filters. A large number of the stars had data available in both the $g$ and $r$ filters. This allowed us to calculate the $T_{\rm eff}$ of a given star from their difference using the relation (\ref{Teff}) \citep{Fukugita2011-bc}. This empirical relation provides an approximation of $T_{\rm eff}$ and should be interpreted with caution for highly evolved cool giants such as Mira variables.

\begin{align}
T_{\rm eff} = \frac{1.09}{(g-r)+1.47} \cdot 10^4 [K]\label{Teff}
\end{align}

The mjd, magnitude, and magnitude uncertainty of each light curve were used to independently verify the catalogue classification using UPSILoN (AUtomated Classification of Periodic Variable Stars using MachIne LearNing; \citealt{Kim2016}). This is a Python-based program that can classify periodic variable stars, which includes Miras, based on Machine learning. It needs mjd, mag and mag error as input data regardless of the filter chosen. As output, for each star that contains more than 80 data points, it gives us the variable type of the star, the probability with which it is sure that it is the variable type, and whether the classification was successful or the classification is suspicious -- period is in period alias and/or period is very low. For our stars in the galactic halo, the algorithm classified that most of our stars are indeed Miras, and only 19 of them fall into a different category, namely the semi-regular variables (SRV) category, where this type is often confused with Miras. For stars in the galactic bulge, the program did not have enough points for most stars, so it determined the type of variability with very low probability. Still, for those it was sure of, it determined them as Miras.

From the obtained $g-r$ and $r$ we could then calculate the values for the unredening color $(g-r)_0$, and also the absolute magnitude in the $r$ filter ($M_{\rm r}$) for each of our stars. The following relations were used for the calculation: Eqs. (\ref{Mr}) and  (\ref{(g-r)0}), where $d$ is the distance in parsecs, $A_{\rm r}$ is the extinction in $r$ (mag), and $E(g-r)$ is the reddening, also called color extinction. 

\begin{align}
M_{\rm r} = r-5\log(d)+5-A_{\rm r} \label{Mr}
\end{align}

\begin{align}
(g-r)_0 = (g-r)-E(g-r) \label{(g-r)0}
\end{align}

Extinction corrections were applied before calculating intrinsic colors and absolute magnitudes. To calculate $(g-r)_0$ and $M_{\rm r}$, however, we lacked knowledge of the magnitude of $A_{\rm r}$ and $E(g-r)$ for Miras. Since the ZTF is calibrated to Pan-STARRS1, see \citep{Ngeow2021-tt} and \citep{2017yCat.2349....0C}, we were able to use the conversion relations associated with it since the ZTF does not have its own transformation relations between filters. Extinction values were obtained in the Johnson–Cousins system and transformed to the Pan-STARRS1 photometric system. We also obtained values for $A_{\rm v}$ and $E(B-V)$ using the IRSA DUST archive \citep{35}. These values were then converted to the $g$ and $r$ filters using the following Eqs: (\ref{Ar}), (\ref{Ag}) and (\ref{E(g-r)}). 

\begin{align}
A_{\rm r} = 0.843\,A_{\rm v} \label{Ar} \\
A_{\rm g} = 1.155\,A_{\rm v}\label{Ag} \\
E(g-r) = \frac{A_{\rm r}}{2.617} \label{E(g-r)}
\end{align}

The values of 0.843 and 1.155 were taken from the transformation relations given in \cite{2019ApJ...877..116W}. The value of 2.617 corresponds to the quantity $R_{\rm r}$, which is the extinction coefficient for the $r$ filter, whose value is constant for a given color \citep{Ngeow2021-tt}.

\subsection{Obtaining quantities from isochrones}

The calculated $M_{\rm r}$ and $(g-r)_0$ values for each star were then used as input values to the Stellar Isochrone Fitting Tool - StiFT \citep{2021osvm.confE...1S} in txt format. The PARSEC isochrone grid \citep{2012MNRAS.427..127B} in the Pan-STARRS1 photometric system was adopted as the reference grid for the interpolation. The observed values of $M_r$ and $(g-r)_0$ were used as input parameters, while stellar age, $T_{\rm eff}$, luminosity, and mass were obtained by interpolation within the isochrone grid. Stellar parameters were estimated by interpolating within the adopted isochrone grid. 

As our input data, we used 88 Miras in the bulge (only for this number do we have values in the $g$ and $r$ filters) and 5 for clusters in the bulge. We got results for 81 and 2 Miras in the bulge, respectively. For the 348 Miras found in the halo, we had data in both $g$ and $r$ filters, including 1 in the cluster. We successfully counted data for 334 stars in the halo.

\subsection{Extinction and reddening values} \label{reddening}

 Interstellar extinction is particularly significant towards the Galactic bulge and therefore must be taken into account when deriving intrinsic stellar parameters. In contrast, extinction towards the Galactic halo is generally much lower. In many studies, halo extinction is considered negligible. Both reddened and dereddened quantities were analysed and compared throughout this work.

The values of extinction and reddening in the $V$ and $I$ filters ($E(V-I)$, $A_{\rm I}$) were obtained from OGLE-III \citep{2009IAUS..256...30S}, which provides extinction and reddening estimates. The optical gravitational lensing experiment (OGLE) also provided values for the extinction ratio between the $V$, $I$ and $J$, $K$ filters ($R_{\rm VIJK}$), which allowed us to recalculate the extinction and reddening from the $V$ and $I$ filters to the $J$ and $K$ filters. We used the classical conversion (\ref{EJK}) and (\ref{AK}), where the constants 0.36 and 1.36 were taken from \citep{2019AJ....158..105L}. 

\begin{align}
E(J-K) = E(V-I)\,R_{\rm VIJK} \label{EJK}
\end{align}

\begin{align}
A_{\rm K} = \frac{0.36\,E(V-I)}{1.36} \label{AK}
\end{align}

 We also used the Gaia DR3 VariSumarry catalogue, which provided us with their extinction and reddening values in addition to the $G$, $BP$, and $RP$ colors, both for the mean values and for the values at the minimum and maximum of the stellar brightness during the cycle. 
 
To minimise the effects of interstellar extinction, Wesenheit magnitudes were also calculated \citep{1976RGOB..182..153M}. Wesenheit magnitudes combine photometric measurements from two filters to produce quantities that are largely insensitive to reddening. 

The Wesenheit magnitude was calculated by combining photometric measurements from two filters using the extinction coefficient $R$, defined as the ratio of total to selective extinction \citep{2005MNRAS.360.1033N}. For the $V$ and $I$ bands, the Wesenheit magnitude was calculated using Eqs.~(\ref{WI}) and (\ref{RVI}), adopting $R=1.55$ following \citet{2022AJ....164..154N}. An analogous relation was used to derive the reddening-free magnitude in the $K$ band, as given in Eq.~(\ref{WK}). The ZTF $g$, $r$, and $i$ photometry was transformed to the Pan-STARRS1 system using the relations described above and subsequently used to calculate the corresponding Wesenheit magnitudes from Eqs.~(\ref{Wgi})--(\ref{Wri}). Absolute Wesenheit magnitudes were then derived using Eq.~(\ref{M_Wgr}) following \citet{Ngeow2021-tt}.

\begin{align}
W_{\rm I} = I-R(V-I) \label{WI} \\
R=A_{\rm I}/E(V-I) \label{RVI} \\
W_{\rm K} = K-0.618(J-K) \label{WK} \\
W_{\rm gi} = g - 2.274(g-i) \label{Wgi} \\
W_{\rm gr} = r - 2.905(g-r) \label{Wgr} \\
W_{\rm ri} = r - 4.051(r-i) \label{Wri} \\
M_{\rm Wgr} = W_{\rm gr} - 5\log(d) + 5 \label{M_Wgr}
\end{align}

\section{Analysis}
All the data were processed with custom code. These results will be commented on in the following sections, first for the galactic bulge and then for the galactic halo.
\subsection{Galactic bulge}

\subsubsection{Metallicity and period distribution of Miras} \label{metalperiod}

\begin{figure} [h]
\centerline{\includegraphics[width=8cm]{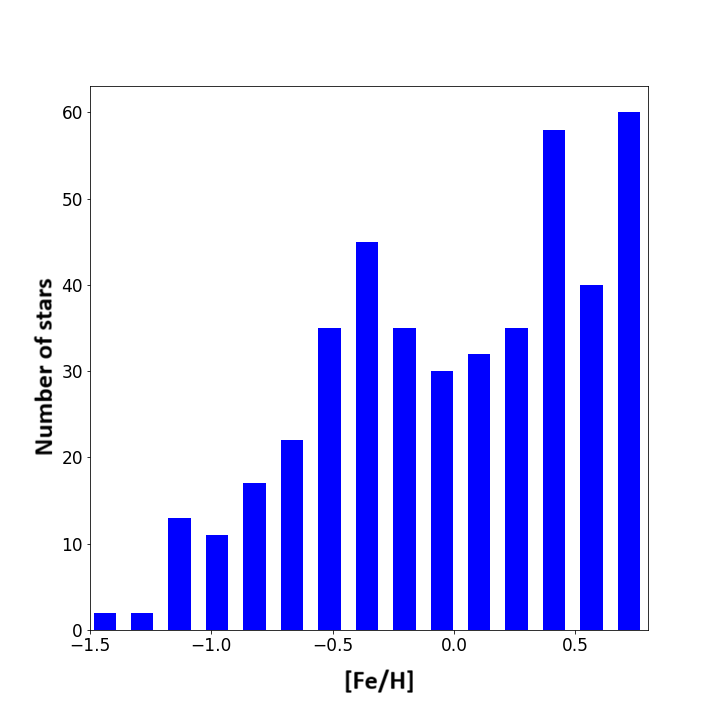}}
\caption {Abundance rate of metallicity. There is a noticeable bimodal distribution for Miras in the galactic bulge with peaks of \text{--}0.4 dex and $+$0.5 dex.} 
\label{obr.histFe}
\end{figure}

Figure~\ref{obr.histFe} shows the metallicity distribution of Mira variables in the Galactic bulge. The distribution exhibits two prominent peaks at approximately $\mathrm{[Fe/H]}$ $-0.4$ dex and $+0.5$ dex, indicating a bimodal metallicity distribution. A similar bimodality has been reported by \citet{RojasArriagada2019} and is generally interpreted as evidence for two stellar populations in the Galactic bulge. The metal-rich component is associated with the secular evolution of the Galactic bar, whereas the origin of the metal-poor component remains uncertain and has been linked to either the early thick disc or an early phase of bulge formation.

\begin{figure} [h!!]
\centerline{\includegraphics[width=8cm]{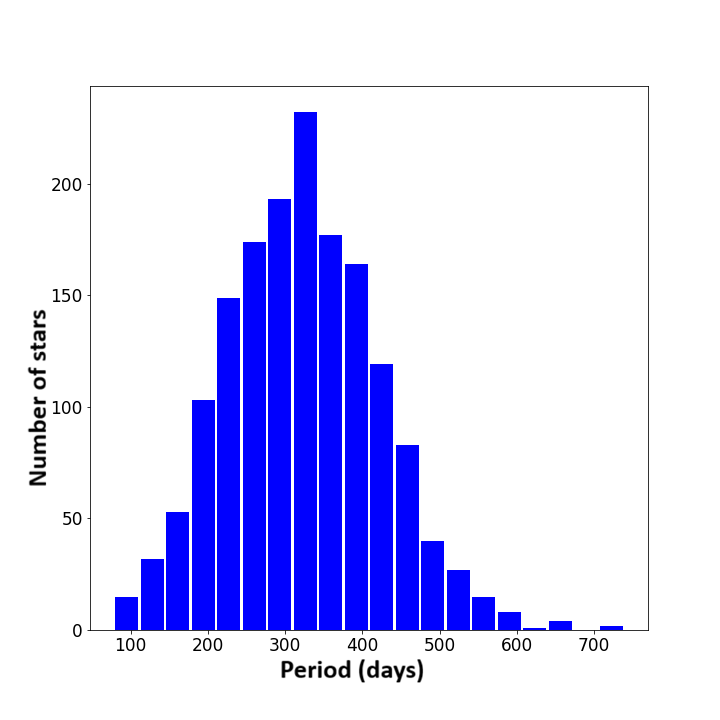}}
\caption{Dependence of the number of stars on period showing a Gaussian distribution with the peak around 330 days.} 
\label{obr.histPer}
\end{figure}

The period distribution shown in Fig.~\ref{obr.histPer} is approximately Gaussian and peaks at about 330 days. This value is consistent with the typical pulsation periods of Mira variables and agrees well with the distribution reported by \citet{Matsunaga2017} for bulge Miras identified from the OGLE III survey. Despite using different catalogues, both studies show a remarkably similar period distribution. 

\subsubsection{Color\text{--}magnitude and color\text{--}color relations} \label{color-magnitude}

\begin{figure} [h!!]
\centerline{\includegraphics[width=8cm]{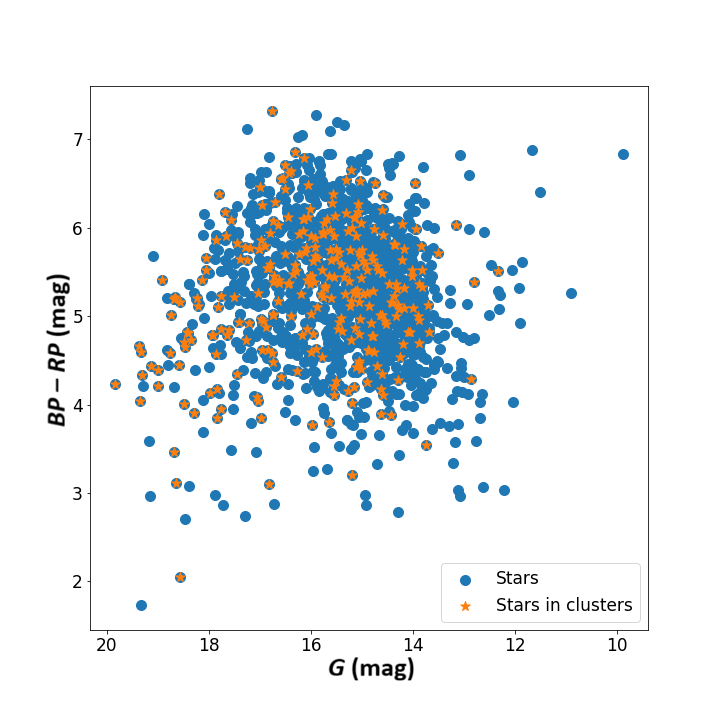}}
\caption{Gaia color magnitudes used for Miras stars and Miras located in clusters.} 
\label{fig:colorGaia_a}
\end{figure}

From the Gaia catalogue, for the $G$, $BP$, and $RP$ magnitudes, we obtained Fig. \ref{fig:colorGaia_a}. The figure shows that the Mira variables occupy the expected region of the CMD corresponding to luminous, cool AGB stars. Increasing $BP-RP$ values correspond to cooler stellar temperatures, as expected for Mira variables.
The observed scatter is likely caused by a combination of interstellar reddening, photometric uncertainties, and the intrinsic variability of Mira stars. Another reason is that for $BP-RP$ values $BP-RP <$ 6.5 mag, we have crossed the threshold below which the Gaia values are not taken into account because they are unreliable. This limit can also be found for the $G$ filter, where the limit is 16 mag and values above this magnitude are unreliable too, \citep{Mowlavi2018}. In our plot, we can also spot these data, and they are located in the lower left part of the plot and form a prominent part of an otherwise homogeneous cluster.

\begin{figure} [h!!]
\centerline{\includegraphics[width=8cm]{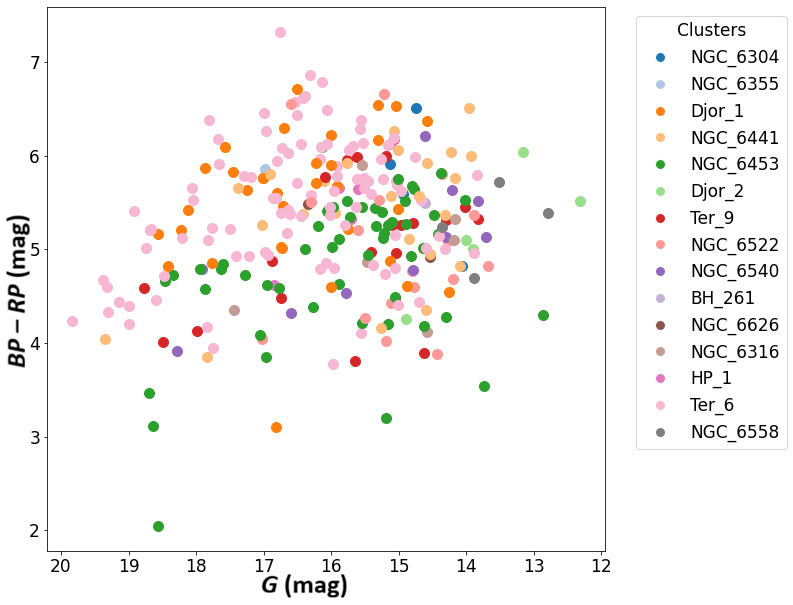}}
\caption{Gaia color magnitudes used for color\text{--}magnitude diagram with location of Miras belonging to specific clusters.} 
\label{fig:colorGaia_b}
\end{figure}

If we look at Fig. \ref{fig:colorGaia_b}, where we can see the distribution of the stars in the clusters, it is clear that Ter 6 and Djor 1 are located in areas with a higher color index, so they appear redder than the stars in NGC 6453 and NGC 6441, which are located below them.  This is due to the previously mentioned reddening effect, where, according to \cite{Bica2016}, the Ter 6 cluster has a higher reddening than NGC 6441. The reddening values for Djor 1 and NGC 6453 were taken from \cite{Harris1976} and show the same reddening trend with respect to each other.

\begin{figure} [h!!]
\centerline{\includegraphics[width=8cm]{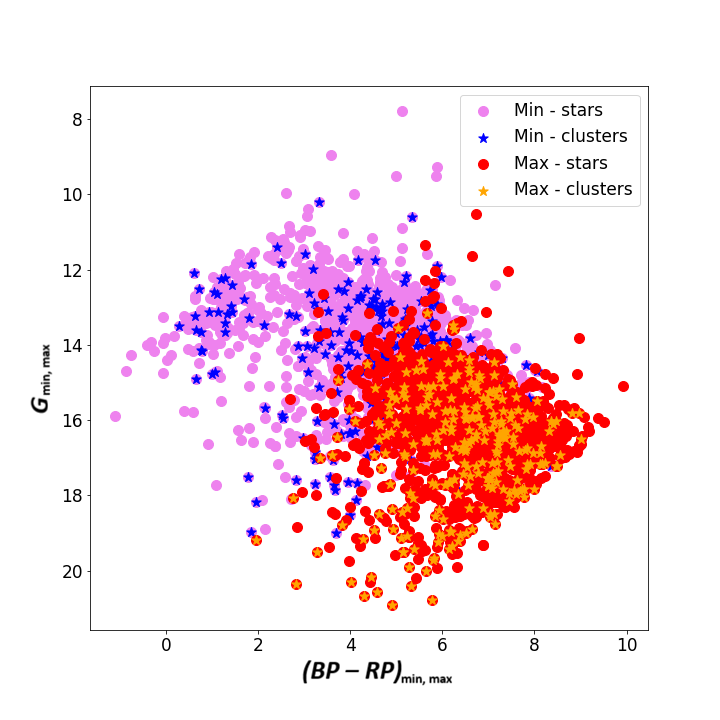}}
\caption{Shift of Miras in color\text{--}magnitude diagram during their pulsation cycle.} 
\label{obr.colorgaiaminmax}
\end{figure}

For Gaia, in addition to the mean values for $G$, $BP$, and $RP$, we also obtained their minimum and maximum values, which are shown in Fig. \ref{obr.colorgaiaminmax}. The minimum-light and maximum-light positions reflect the expected color and luminosity changes during the pulsation cycle, with Mira variables becoming brighter and bluer near maximum light.

\begin{figure} [h!!]
\centerline{\includegraphics[width=8cm]{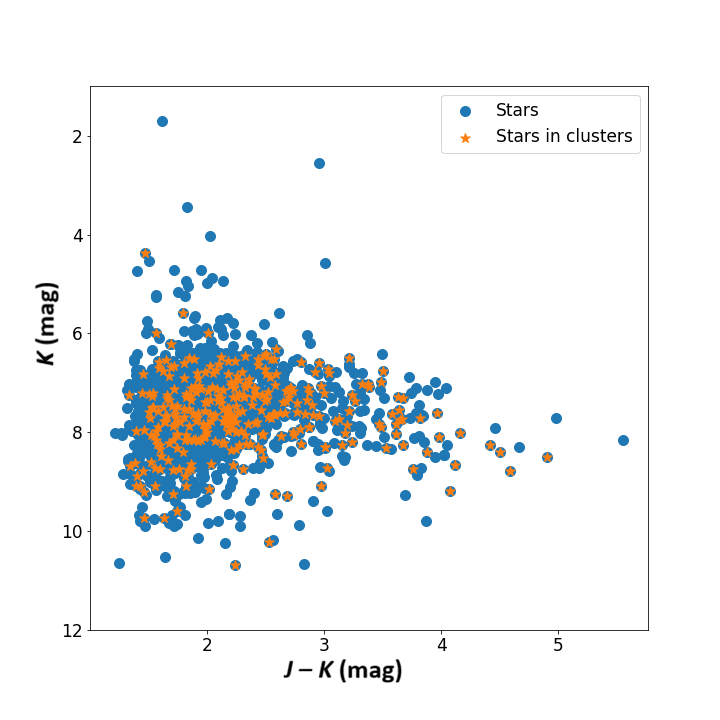}}
\caption{Location of Miras and Miras in clusters in color\text{--}magnitude diagram for 2MASS filters.} 
\label{obr.colorJK}
\end{figure}

Since Miras can be divided into oxygen-rich (O-rich) and carbon-rich (C-rich) stars, their chemical type can be estimated from the dependence of the $J-K$ color on the $K$ magnitude. Carbon-rich Miras exhibit strong CN and C$_2$ absorption bands, resulting in redder $J-K$ colors, whereas O-rich Miras are dominated by TiO and H$_2$O absorption, which affects both the $J$ and $K$ bands \citep{Matsunaga2017}. Using observations of Miras in the Large Magellanic Cloud (LMC), \citet{Matsunaga2017} proposed empirical limits of $K < 9.5$ mag and $J-K > 3.2$ mag to identify C-rich Miras. Applying these criteria to our sample, approximately 40 Miras satisfy these criteria (Fig.~\ref{obr.colorJK}).

\begin{figure} [h!!]
\centerline{\includegraphics[width=8cm]{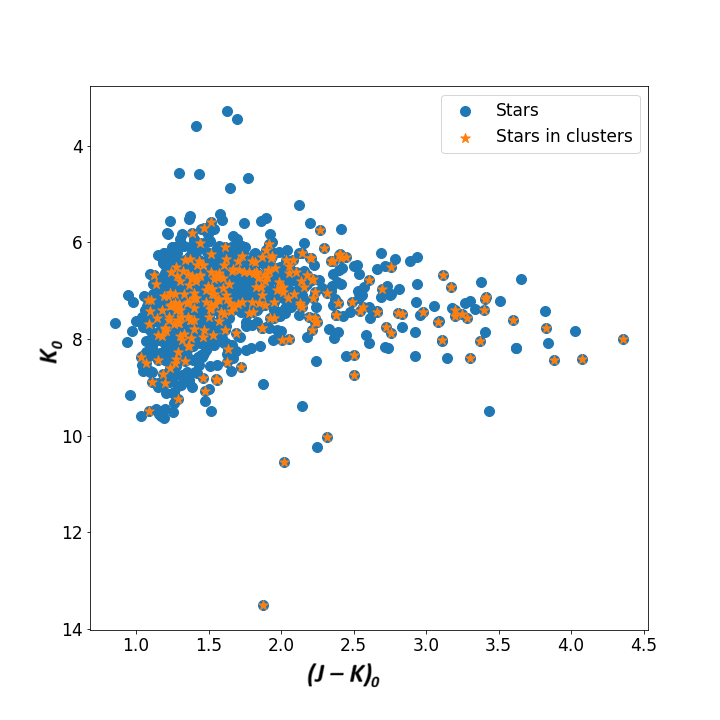}}
\caption{Dereddened 2MASS magnitudes for bulge Miras.} 
\label{obr.bulgereddJK0}
\end{figure}

Just as we examined the color\text{--}magnitude diagram for the $J$ and $K$ filters in Fig. \ref{obr.colorJK}, because of the interstellar dust and gas, it was necessary to obtain values for extinction and reddening in these filters as well. From this, we were then able to convert these values to dereddened values of $(J-K)_0$ and $K_0$, the resulting dependence of which is shown in Fig. \ref{obr.bulgereddJK0}. When we include the reddening, most stars shift towards bluer colors after the reddening correction. Thus, the interstellar material did indeed make the stars slightly redder than their true color. Only a relatively small shift is observed along the magnitude axis.

\begin{figure} [h!!]
\centerline{\includegraphics[width=8cm]{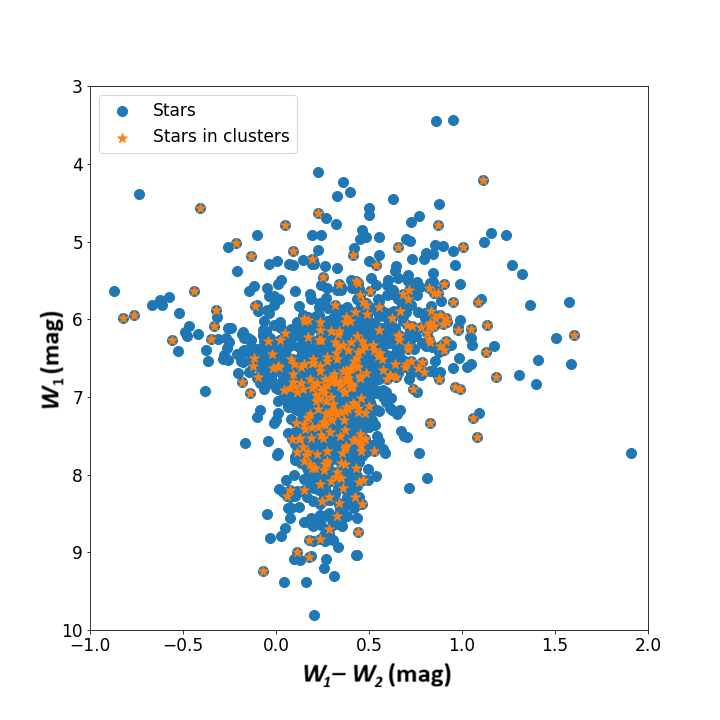}}
\caption{Location of Miras and Miras in clusters in color\text{--}magnitude diagram for filters $W_1$ and $W_2$ in WISE catalogue. These filters belong to the near-infrared part of the spectra.} 
\label{obr.colorW12}
\end{figure}

Semi-regular (SR) variables are frequently misclassified as Miras because of their similar photometric properties, although they generally exhibit smaller amplitudes and shorter periods. Following the criterion proposed by \citet{Chen2020}, stars with $W_1-W_2 \gtrsim 0.4$ are likely to be SR variables. Applying this limit to our sample (Fig.~\ref{obr.colorW12}) places approximately one third of the stars in the SR region. However, the authors emphasise that the overlap between Miras and SR variables is considerable, and that variability amplitude, unavailable for our sample, provides a more reliable classification. Consequently, this separation should be regarded only as tentative.

\begin{figure} [h!!]
\centerline{\includegraphics[width=8cm]{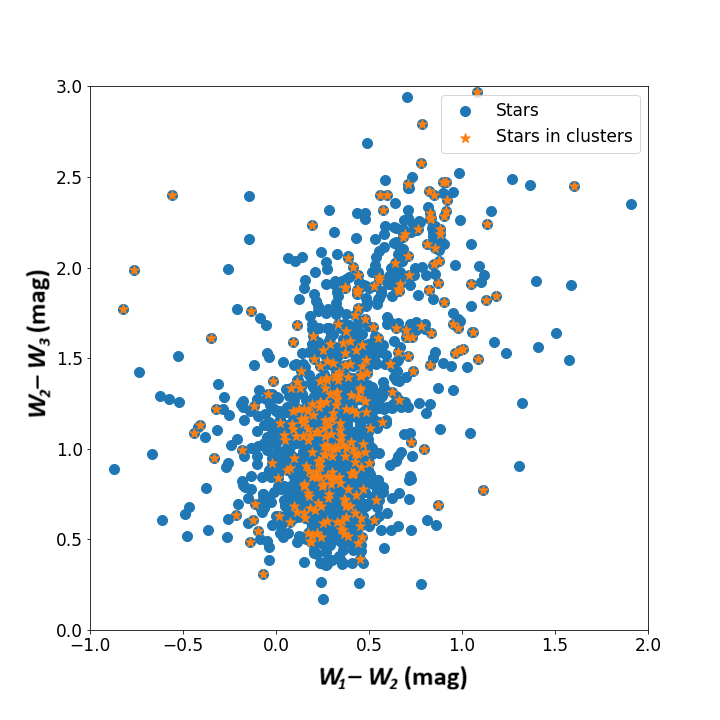}}
\caption{Location of Miras and Miras in clusters in color\text{--}color diagram for filters from WISE catalogue.To note, there is an extended linear dependence, whereas the color $W_1 - W_2$ grows, so does the color $W_2 - W_3$.} 
\label{obr.colorW123}
\end{figure}

A more informative classification is provided by the $W_1-W_2$ versus $W_2-W_3$ color diagram (Fig.~\ref{obr.colorW123}). Both \citet{Chen2020} and \citet{Iwanek2021} identify stars with $-0.2 < W_1-W_2 < 0.25$ and $W_2-W_3 > 0$ as O-rich Miras. However, they disagree on the interpretation of the region defined by $0.25 < W_1-W_2 < 1.5$, where \citet{Chen2020} classify the stars as SR variables, whereas \citet{Iwanek2021} identify them as C-rich Miras. Since the latter study is based exclusively on LMC stars, this discrepancy may reflect differences between stellar populations and should be considered when applying these criteria to Galactic bulge objects.

One of the most widely used methods for separating O-rich and C-rich Miras combines Gaia and 2MASS photometry through Wesenheit magnitudes, which largely remove the effects of interstellar reddening. This approach is particularly suitable for Galactic bulge stars, where reddening and source confusion make classifications based solely on 2MASS colors less reliable than in the Magellanic Clouds.

\begin{figure} [h!!]
\centerline{\includegraphics[width=8cm]{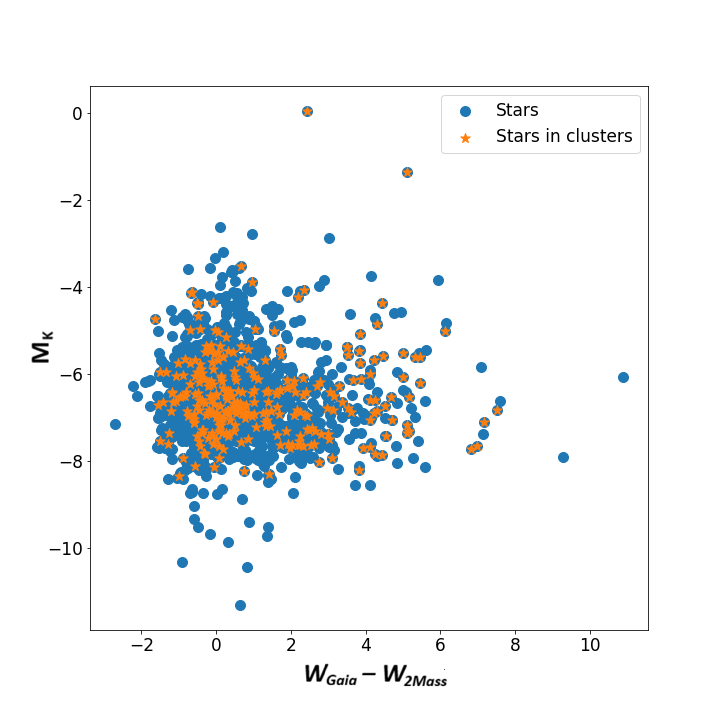}}
\caption{Comparing color with absolute magnitude from two different catalogues - 2MASS and Gaia. } 
\label{obr.colorWgk}
\end{figure}

Following the empirical boundary derived by \citet{Sun2023}, which separates O-rich and C-rich Miras in the Gaia--2MASS Wesenheit diagram, we classified our sample using the diagram shown in Fig.~\ref{obr.colorWgk}. The boundary has reached a minimum color of approximately 2.6 mag at $M_K \approx -7.5$ mag, reflecting differences in molecular absorption between O-rich and C-rich atmospheres. Applying this criterion to our data indicates that approximately one third of the sample belongs to the C-rich Mira population.

\begin{figure} [h!!]
\centerline{\includegraphics[width=8cm]{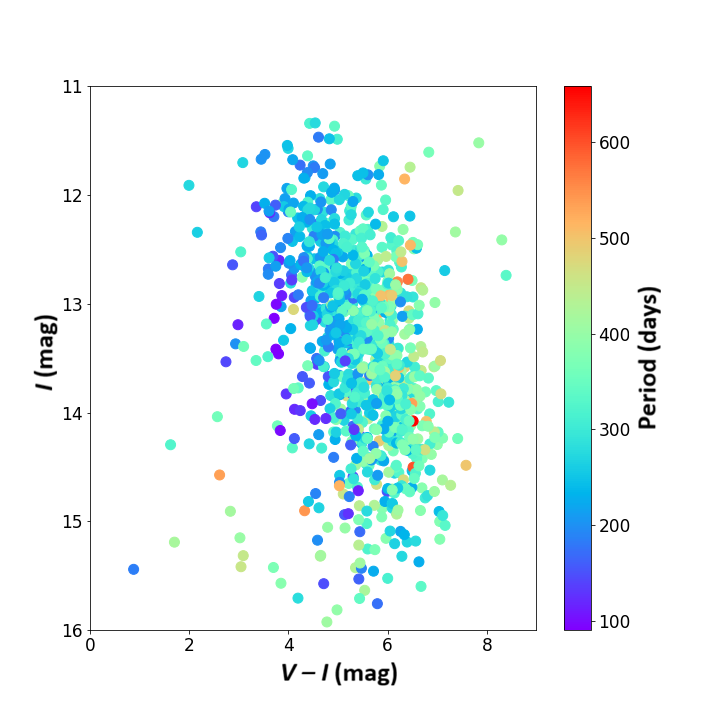}}
\caption{Comparing Miras in color\text{--}magnitude diagram for OGLE III filters. It is clear that there is color dependence on period, which changes with higher values of color.} 
\label{obr.colorVI}
\end{figure}

Figure~\ref{obr.colorVI} shows the color-magnitude diagram based on the OGLE III $V$ and $I$ filters. The distribution of the stars is consistent with the expected AGB population and exhibits a clear correlation with the pulsation period: longer-period Miras are brighter and redder. This behaviour agrees with the well-established period--luminosity and period-color relations of Mira variables.

\begin{figure} [h!!]
\centerline{\includegraphics[width=10cm]{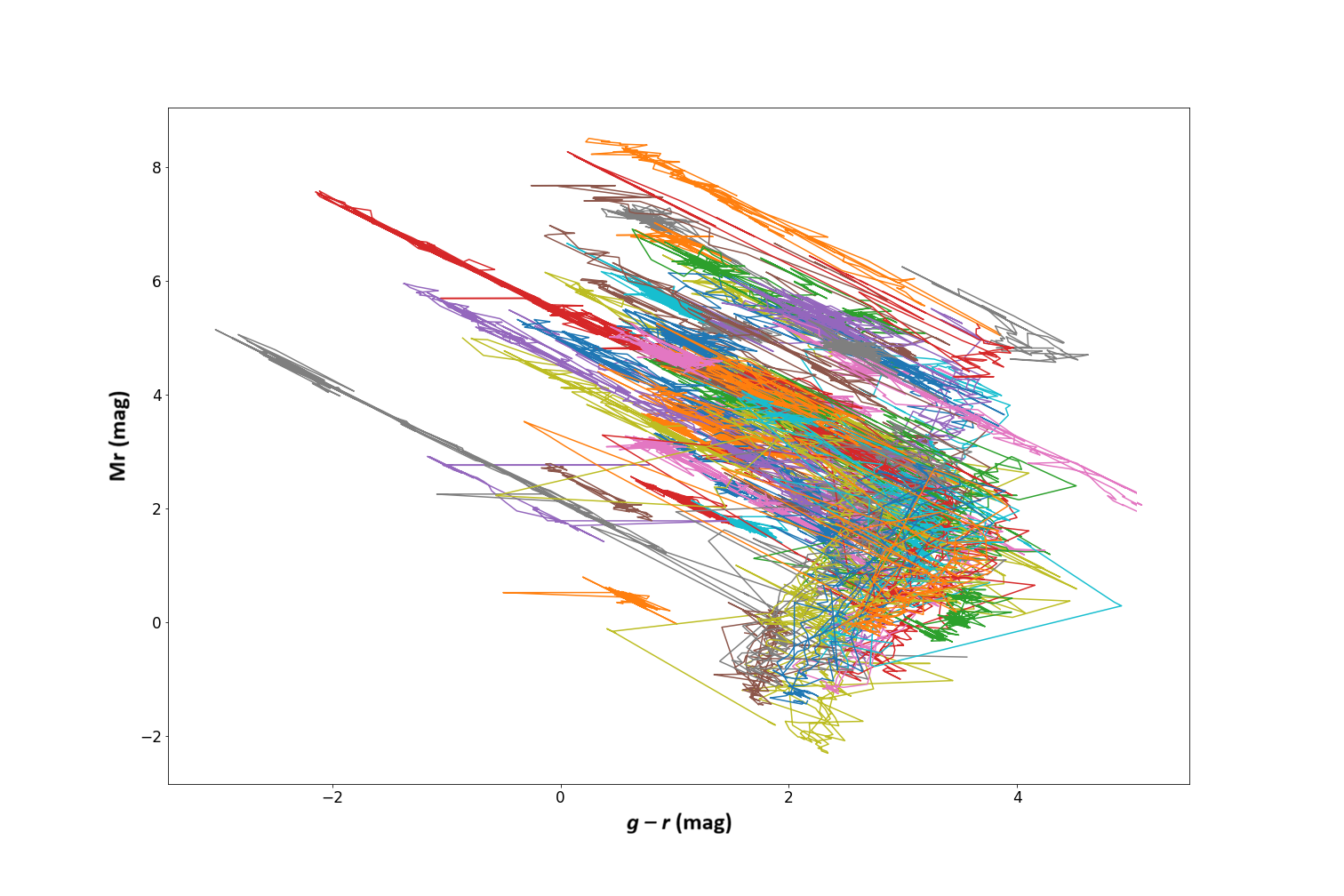}}
\caption{Moving CMD for bulge Miras. Different colors indicate different Mira stars.} 
\label{obr.movingCMD}
\end{figure}

A moving CMD was constructed, see Fig. \ref{obr.movingCMD}, from the data obtained from the ZTF catalogue, as discussed in the methodology section. The linear dependence of color on absolute magnitude remains the known trend using Gaia bands, except that the trend is reversed. Although both Gaia and ZTF observe in the optical wavelength range, their filters differ substantially in bandwidth and spectral coverage. Gaia employs very broad $G$, $BP$, and $RP$ passbands, whereas ZTF uses narrower Sloan-like $g$ and $r$ filters. Since Mira spectra are dominated by strong phase-dependent molecular absorption bands, particularly TiO and VO in O-rich stars and CN and C$_2$ in C-rich stars, changes in $T_{\rm eff}$ and molecular opacity affect the Gaia and ZTF passbands differently during the pulsation cycle. In addition, a minor contribution from circumstellar dust emission at the longest wavelengths may further modify the observed color variations. These effects can produce different color-magnitude trajectories in the two photometric systems.

\subsubsection{Period magnitude relations}

\begin{figure} [h!!]
\centerline{\includegraphics[width=8cm]{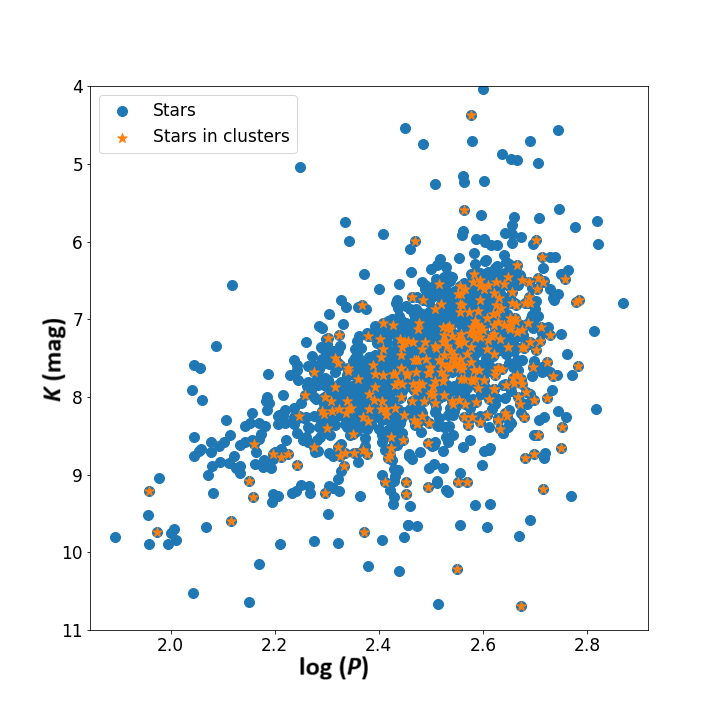}}
\caption{Period dependence on 2MASS filter log($P$) on $W_1$.} 
\label{obr.logp_k}
\end{figure}

Figure~\ref{obr.logp_k} shows the period--luminosity relation between $\log P$ and the $K$-band magnitude. The relation broadens for $\log P \approx 2.4$--2.8, where we have multiple magnitudes for a given period. This increased scatter is consistent with the presence of C-rich Mira variables, which are known to exhibit a larger dispersion in the period-luminosity relation. A similar behaviour was reported in \cite{https://doi.org/10.48550/arxiv.astro-ph/0512578} and \cite{Yuan2017}, although for Miras located in the LMC and SMC.

\begin{figure} [h!!]
\centerline{\includegraphics[width=8cm]{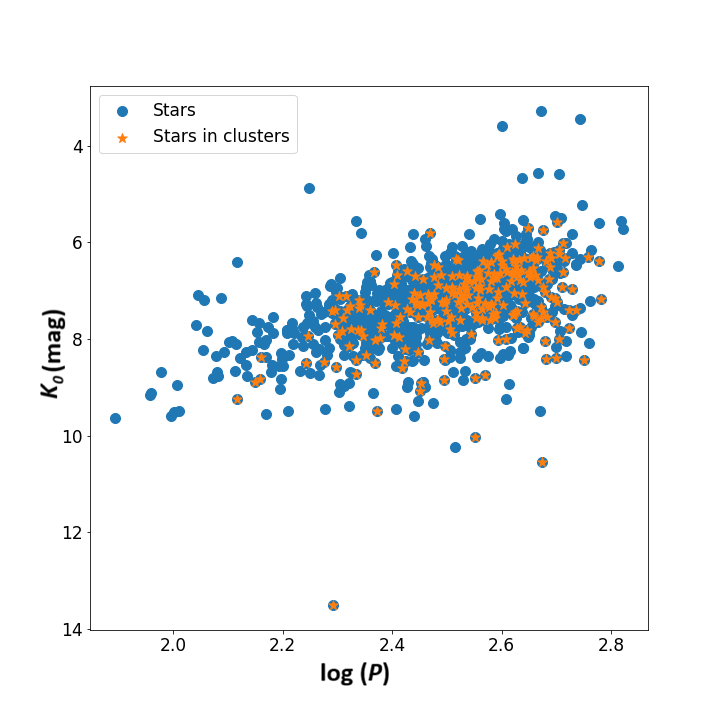}}
\caption{Period dependence on dereddened $K_0$ filter.} 
\label{obr.bulgereddPJK0}
\end{figure}

The same relation is shown in Fig. \ref{obr.bulgereddPJK0}, this time using dereddened values of $K_0$. The dereddened relation exhibits a smaller dispersion in the $K_0$ magnitudes, indicating that interstellar extinction contributes significantly to the observed scatter in the original diagram. 

\begin{figure} [h!!]
\centerline{\includegraphics[width=8cm]{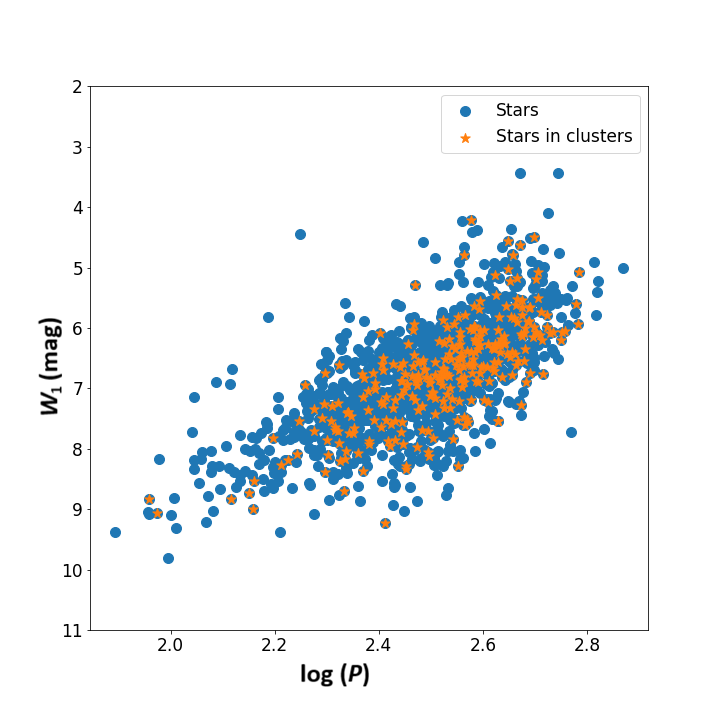}}
\caption{Linear dependence of log($P$) on $W_1$ for Miras and Miras in bulge globular clusters.} 
\label{obr.logp_w1}
\end{figure}

The same dependence can be seen in Fig. \ref{obr.logp_w1}, which shows the dependence of log$P$ on $W_1$, but with the difference that the relation is noticeably tighter and clustered together, compared to the previous graph. Our graph is quite comparable to the graph obtained in the paper by \cite{Schultheis2008}. There may be several reasons why we observe values closer together. The reduced scatter is expected because the $W_1$ band is less affected by interstellar extinction than the $K$ band and is also less sensitive to temperature variations during the pulsation cycle. This behaviour is consistent with the well-known reduction of scatter in the period--luminosity relation towards longer wavelengths.

\subsection{Analysis of galactic halo}
\subsubsection{Metallicity and period distribution of Miras}

\begin{figure} [h!!]
\centerline{\includegraphics[width=8cm]{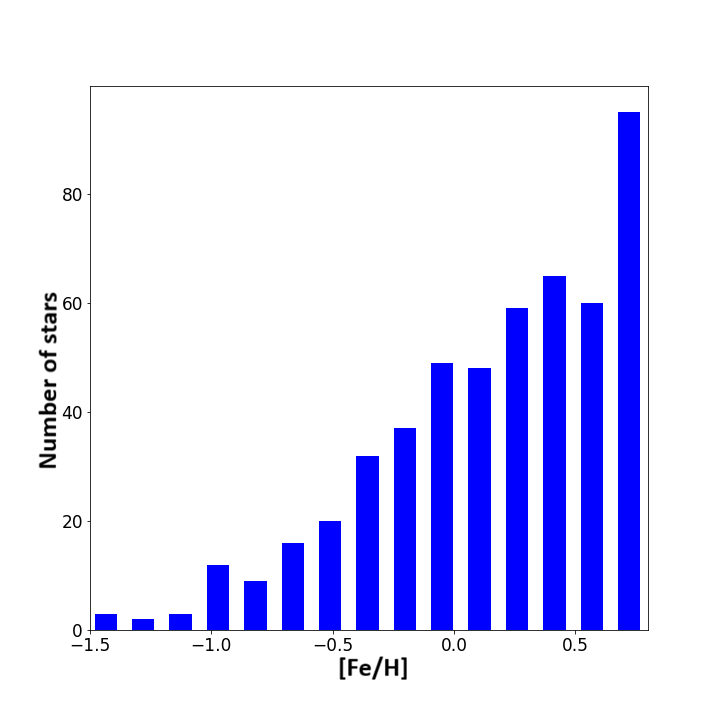}}
\caption{Distribution of metallicities for halo Mira variables.} 
\label{obr.halohistfe}
\end{figure}

The distribution of metallicities by frequency does not have a bimodal distribution for the halo, as can be seen in Fig. \ref{obr.halohistfe}. The metallicity distribution does not exhibit an obvious bimodal structure, in contrast to the bulge sample. The distribution peaks near $+$0.5 dex, suggesting that these Miras formed in a second phase after the first generations of stars enriched the interstellar medium. 

\begin{figure} [h!!]
\centerline{\includegraphics[width=8cm]{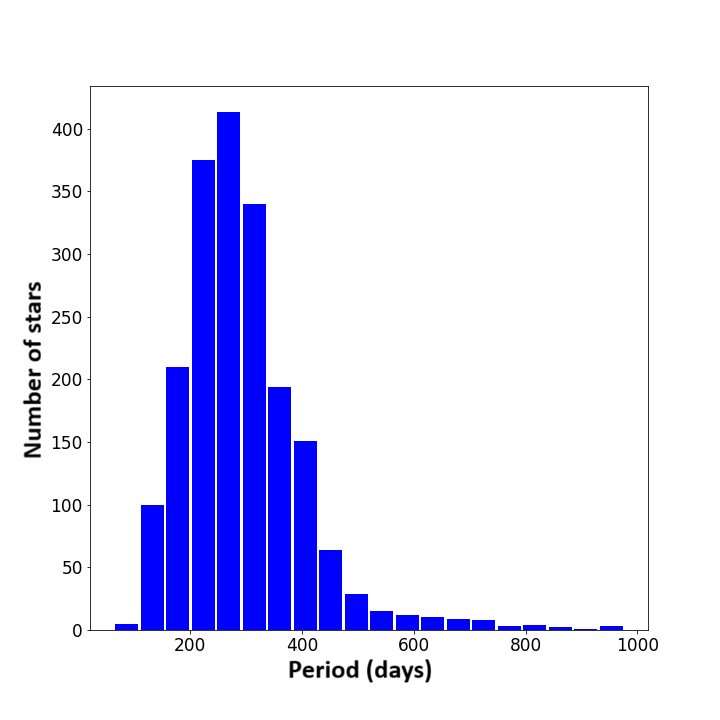}}
\caption{Period distribution of halo Mira variables, peaking at approximately 300 days.} 
\label{obr.halohistper}
\end{figure}

The period distribution shows an approximately Gaussian distribution with a peak for a period of approximately 300 days, see Fig. \ref{obr.halohistper}, which is similar to that found for the bulge sample.

\subsubsection{Color\text{--}magnitude and color\text{--}color relations} 

\begin{figure} [h!!]
\centerline{\includegraphics[width=8cm]{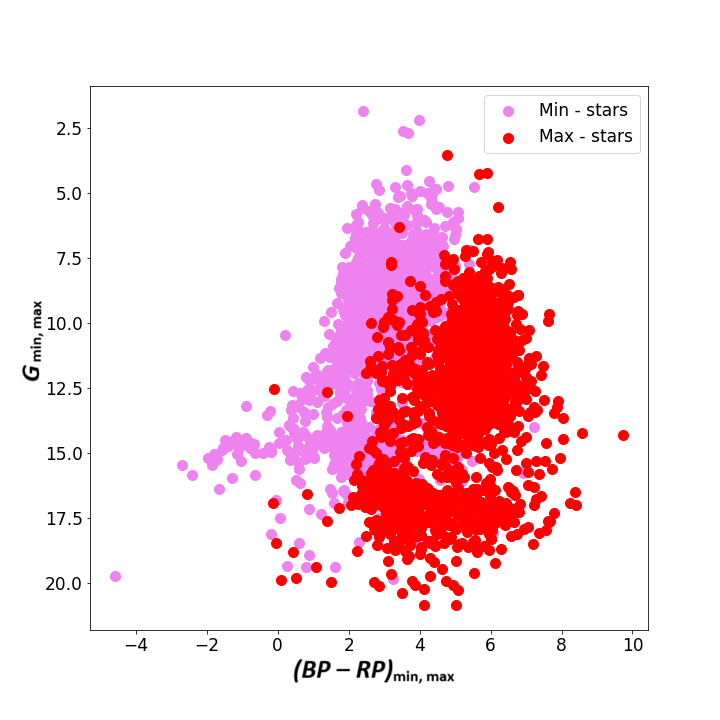}}
\caption{Gaia color-magnitude diagram showing the mean, minimum, and maximum photometric measurements.} 
\label{fig:haloGaia}
\end{figure}

Figure~\ref{fig:haloGaia} shows the Gaia color-magnitude diagram constructed from the $BP$, $RP$, and $G$ photometric bands. The figure also illustrates the displacement of individual stars between their minimum and maximum brightness during the pulsation cycle. The remaining spread in absolute magnitude likely reflects intrinsic differences among the stars together with uncertainties in the distance estimates. Only a relatively small spread in color is observed. This is consistent with the relatively low interstellar extinction towards high Galactic latitudes.

\begin{figure} [h!!]
\centerline{\includegraphics[width=8cm]{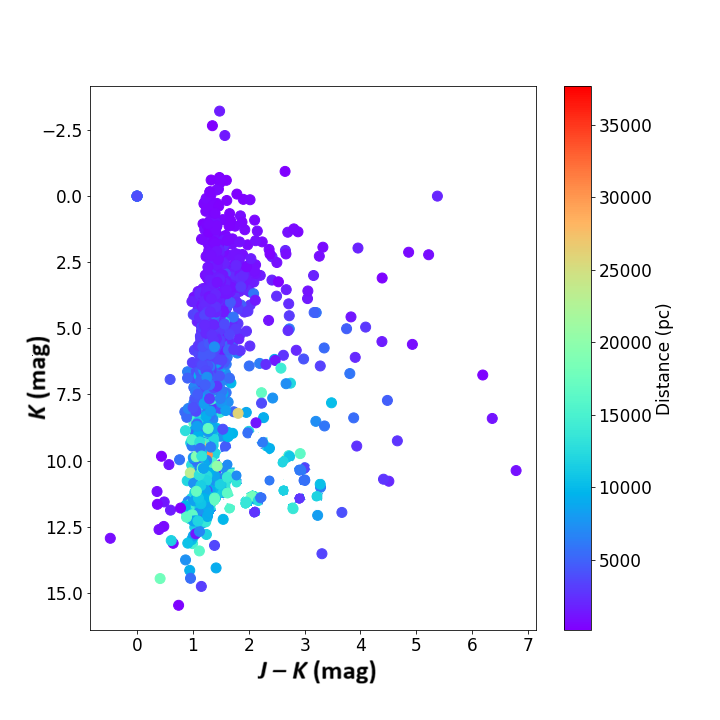}}
\caption{2MASS color\text{--}magnitude diagram.} 
\label{obr.halocolorJK}
\end{figure}

The claim of a wide range of values for magnitude in a particular filter can be verified in Fig. \ref{obr.halocolorJK}, showing filters from the 2MASS catalogue. The wide range of apparent K-band magnitudes is primarily consistent with the large range of distances spanned by the halo sample. This is consistent with the relatively small amount of foreground extinction towards the Galactic halo.

\begin{figure} [h!!]
\centerline{\includegraphics[width=8cm]{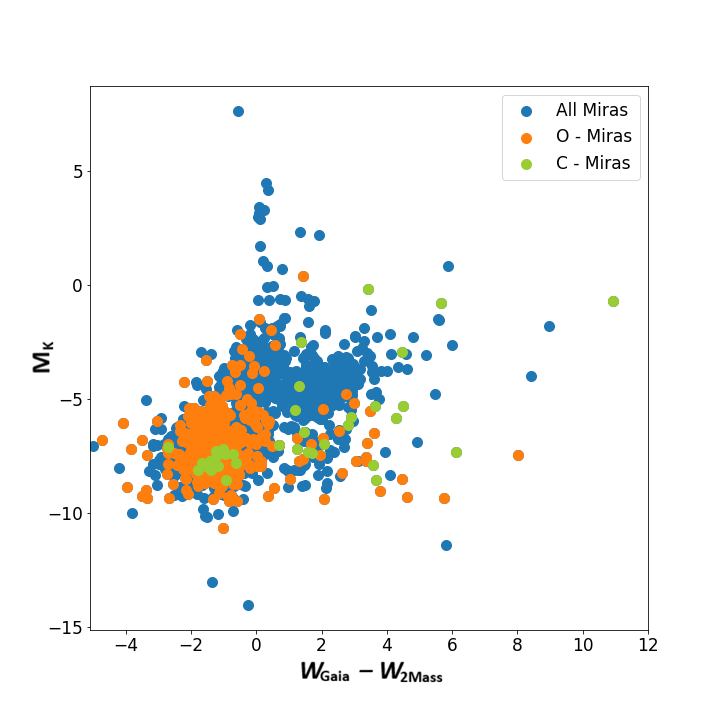}}
\caption{ Gaia--2MASS Wesenheit diagram showing the locations of O-rich and C-rich Mira candidates.} 
\label{obr.halocolorWgjk}
\end{figure}

The UPSILoN classifications were further compared with the photometric separation proposed by \cite{Sun2023}, where, using the difference of the Wesenheit magnitude from the Gaia catalogue and 2MASS versus the absolute magnitude in the $K$ filter, we were able to distinguish C/O Miras, see Fig. \ref{obr.halocolorWgjk}. It should be noted that the C/O Miras determined by us are only indicative based on the light curves, and a more precise determination would require our own spectral analysis and spectroscopy. The light curves obtained by us, based on ZTF data, often have few points or are incomplete. Most O-rich Mira candidates lie within the region defined by Sun et al., although several C-rich candidates also fall inside this region. There is another trend to observe here, namely that C Miras generally show lower absolute magnitudes in the $K$ filter for a given color than O Miras. C-rich candidates tend to occupy lower MK values for a given color. This is consistent with O-rich Miras generally exhibiting bluer near-infrared colors than C-rich Miras,\cite{Feast2006}.

\begin{figure} [t!]
\centerline{\includegraphics[width=11cm]{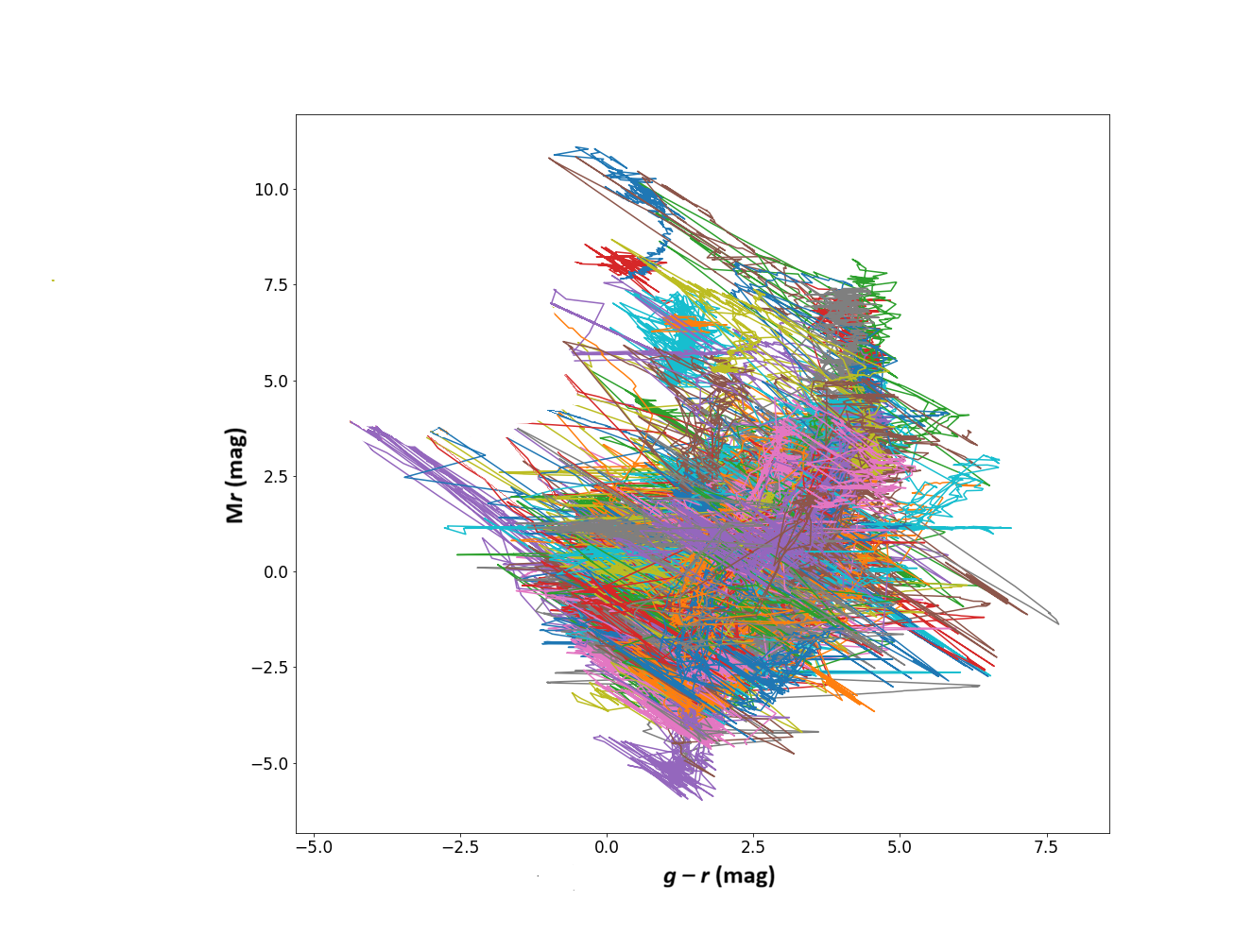}}
\caption{Moving CMD for Miras in the galactic halo. Illustrating the photometric variability of halo Mira variables.} 
\label{obr.haloMCMD}
\end{figure}

The processed moving CMD can be seen in Fig. \ref{obr.haloMCMD}, where we observe the same trend as for the Miras in the bulge. Halo Miras exhibit a wider range of both colors and magnitudes than the bulge sample. The color distribution is shifted towards lower color indices compared to the bulge sample. 

\subsubsection{Period magnitude relations}

\begin{figure} [h!]
\centerline{\includegraphics[width=8cm]{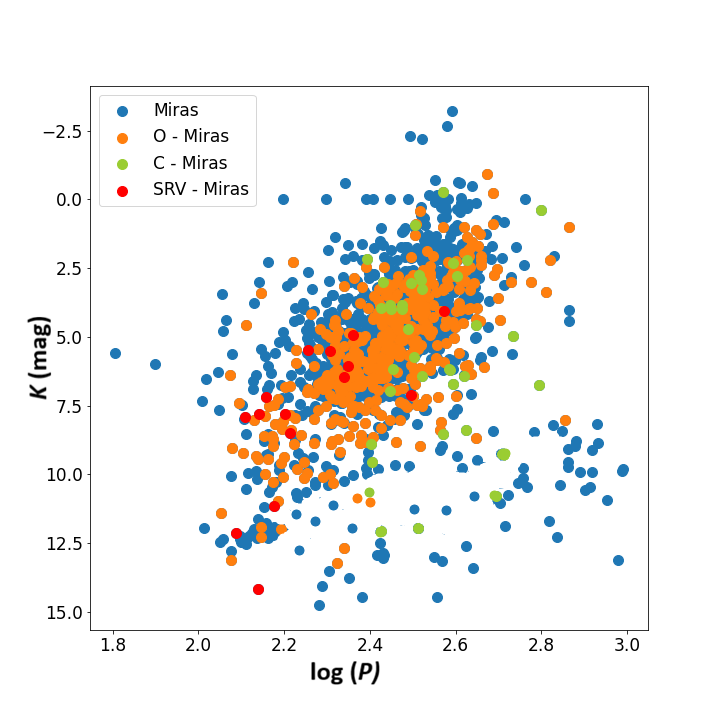}}
\caption{Distribution of O/C Miras and SRV in period\text{--}magnitude diagram.} 
\label{fig:haloperK}
\end{figure}

Figure~\ref{fig:haloperK} shows the period--luminosity relation between $\log P$ and the $K$-band magnitude, and a clear linear trend is observed: Stars with longer pulsation periods are systematically brighter in the $K$ band. In addition, we can see where the O/C Miras and SR variables are located. The graph shows that the C Miras have, on average, longer periods than the O Miras. C-rich Mira candidates tend to occupy the longer-period part of the diagram, consistent with their more advanced evolutionary stage and generally stronger mass loss, \cite{Feast2006}, \cite{Groenewegen1996} and \cite{Groenewegen1995}. However, this separation becomes less distinct at shorter periods.

\begin{figure} [h!!]
\centerline{\includegraphics[width=8cm]{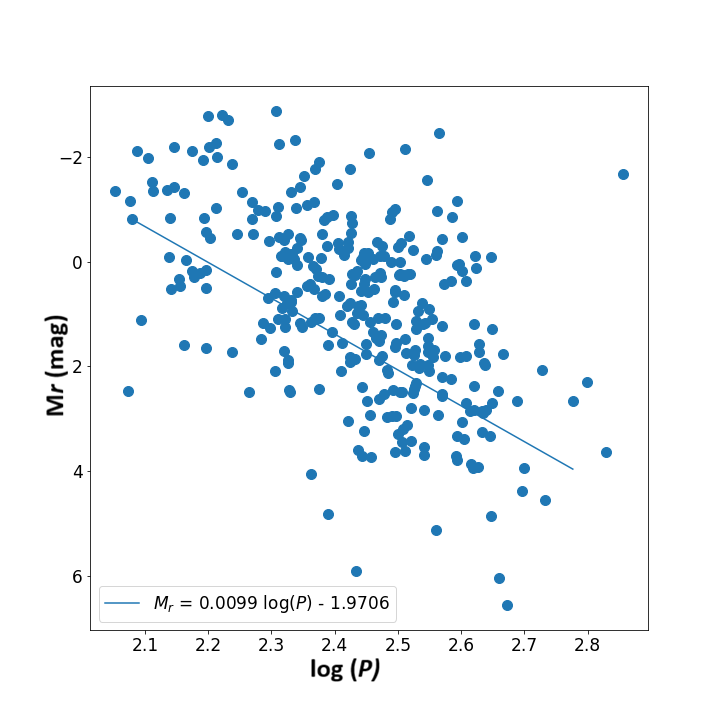}}
\caption{Period\text{--}luminosity diagram for the ZTF $r$ filter shows a linear dependence, which is described by the equation in this figure.} 
\label{obr.halocolorlogPMr}
\end{figure}

The period luminosity (PL) relation for Miras in the halo is shown in Fig. \ref{obr.halocolorlogPMr}. Miras with longer periods are fainter than those with shorter periods. This finding is in agreement with the paper by \cite{Ngeow2023}. Unlike the near-infrared period-luminosity relation, the optical $r$ band exhibits an opposite slope, with longer-period Mira variables appearing fainter. This behaviour is consistent with the increasing influence of molecular absorption and circumstellar extinction in evolved, long-period Mira variables. At the same time, circumstellar dust absorbs optical radiation and re-emits it at infrared wavelengths, reducing the observed optical brightness while enhancing the infrared emission. This is because the $r$ filter is more in the visible part of the spectrum, while $K$, $J$ and $H$ fall in the infrared part of the spectrum. Stars with longer periods exhibit lower observed fluxes in the optical $r$ band than shorter-period Mira variables. These stars have circumstellar envelopes around them that emit in the infrared spectrum, and therefore we observe a decrease in the visible part for longer periods.

The observed period--luminosity relation further supports the use of Mira variables as standard candles to determine distances, especially in older galaxies such as ellipticals, or in spiral galaxy haloes where Cepheids are scarce.

\begin{figure} [h!!]
\centerline{\includegraphics[width=8cm]{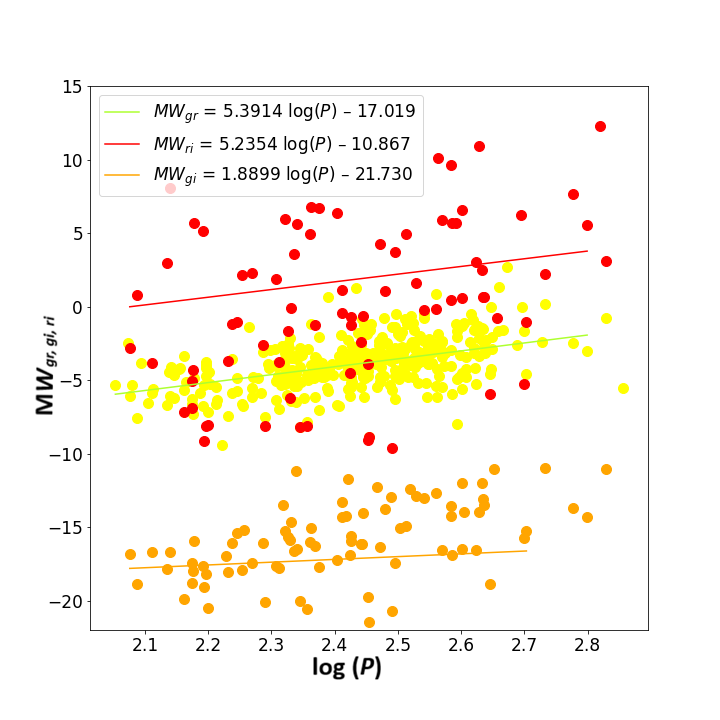}}
\caption{Relation of absolute magnitude in ZTF filters $g$, $r$, and $i$ on log$P$. For each filter, we obtain a similar slope value and linear function.} 
\label{obr.haloreddMWgri}
\end{figure}

Linear period-Wesenheit relations are obtained for all filter combinations, as seen in Fig. \ref{obr.haloreddMWgri}, where for each filter combination we obtained a linear dependence. We have the steepest dependence for the $g-r$ color and the flattest dependence for the $g$, $i$ filters. The different slopes reflect the varying sensitivity of the individual passbands to $T_{\rm eff}$ changes, molecular absorption bands, and circumstellar extinction during the pulsation cycle.

\section{Conclusion}

This study demonstrates that a broad range of astrophysical properties of Mira variables can be investigated using only publicly available multi-survey photometric data, without the need for dedicated spectroscopic observations. By combining Gaia, 2MASS, WISE, OGLE, ZTF, and complementary catalogues, we analysed Mira variables in two different Galactic environments and assessed the diagnostic power of photometric observations for studying their pulsation properties, colors, chemical types, and Galactic distribution.

Our final sample consists of 1608 Mira variables in the Galactic bulge and 1954 in the Galactic halo. Position and proper-motion criteria identified 308 candidate Mira members distributed among 15 bulge globular clusters and four candidates in three halo clusters. Although the cluster memberships of individual stars require spectroscopic confirmation, the adopted photometric and astrometric approach provides an efficient first identification of candidate cluster members in large stellar samples.

The bulge and halo populations exhibit several systematic differences. The bulge sample shows a bimodal metallicity distribution with peaks near $\mathrm{[Fe/H]}\approx-0.4$ dex and $+0.5$ dex, whereas the halo sample displays a single broad maximum near $+0.5$ dex. The period distributions are approximately Gaussian, peaking at about 330 days for the bulge and 300 days for the halo. Photometric classification further suggests that O-rich Mira variables dominate the bulge population, while the halo sample contains a broader mixture of O-rich, C-rich, and semiregular candidates.

The combined use of color-magnitude, color-color, and Wesenheit diagrams proved to be a powerful diagnostic tool. The broader magnitude distribution of halo Mira variables is consistent with their wider distance range, whereas the colors of bulge Miras are more strongly influenced by interstellar extinction. Gaia photometry reproduces the expected displacement of Mira variables during their pulsation cycle, with stars becoming brighter and bluer near maximum light. Among the investigated photometric diagnostics, the Gaia-2MASS Wesenheit diagram provides the most effective separation of O-rich and C-rich Mira candidates, although some overlap between the two populations remains.

The period--magnitude relations obtained from near-infrared photometry clearly show that longer-period Mira variables are systematically brighter and that the relations become progressively tighter towards longer wavelengths. In contrast, the optical ZTF bands exhibit the opposite trend, with long-period Mira variables appearing fainter because of the increasing influence of molecular absorption and circumstellar extinction. Period-Wesenheit relations further reduce the scatter caused by reddening and provide an additional reddening-insensitive diagnostic for studying Mira populations. The moving color--magnitude diagrams derived from ZTF time-series photometry successfully trace the large-amplitude color and brightness variations characteristic of Mira pulsation.

The principal outcome of this work is that combining complementary large-scale photometric surveys provides substantially more information than any individual survey alone. Without requiring new observations, publicly available photometric data allow the investigation of pulsation properties, color evolution, extinction effects, approximate chemical classification, period-luminosity and period-Wesenheit relations, and candidate globular-cluster membership for thousands of Mira variables. While spectroscopy remains essential for confirming chemical abundances and detailed stellar parameters, the present study demonstrates that modern multi-survey photometry alone constitutes a powerful and efficient framework for studying Mira variables across different Galactic environments. These methods are readily applicable to forthcoming large photometric surveys and provide an effective basis for future spectroscopic follow-up and population studies.

\begin{acknowledgements}

This work was carried out within the institutional support framework 
for the development of the
research organization of Masaryk University and was supported by the
grant MUNI/A/1593/2025.

\end{acknowledgements}

\bibliographystyle{aa}
\bibliography{paper.bib}

\end{document}